\def\draftversion{true}

\RequirePackage{ifthen}
\ifthenelse{\equal{\draftversion}{false}}{
\documentclass{nature}
\usepackage{showlabels}
}{
\documentclass{nature}
}

\ifthenelse{\equal{\draftversion}{true}}{
  \marginparwidth 2.7in
  \marginparsep 0.5in
  \newcounter{comm} 
  \def\commnext{\stepcounter{comm}}
  \def\commtext{{\bf\color{magenta}[\arabic{comm}]}}
  \def\commmar{{\bf\color{magenta}[\arabic{comm}]}}
  \def\jim#1{\commnext\marginpar{\small JI\commmar: #1}\commtext}
  \def\jbm#1{\commnext\marginpar{\small JB\commmar: #1}\commtext}
  \def\fgm#1{\commnext\marginpar{\small FG\commmar: #1}\commtext}
  \def\slm#1{\commnext\marginpar{\small SL\commmar: #1}\commtext}
  
}{
  \def\jim#1{}
  \def\jbm#1{}
  \def\fgm#1{}
  \def\slm#1{}
  
}
\usepackage{graphicx}
\makeatletter
\let\saved@includegraphics\includegraphics
\AtBeginDocument{\let\includegraphics\saved@includegraphics}
\renewenvironment*{figure}{\@float{figure}}{\end@float}
\makeatother

\usepackage{soul}  
\usepackage{dcolumn}
\usepackage[colorlinks=true,linkcolor=blue ,citecolor=blue,urlcolor=blue]{hyperref}
\usepackage{multirow}
\usepackage[usenames,dvipsnames]{xcolor}
\usepackage{braket}
\usepackage{bm}
\usepackage{nicefrac}
\usepackage{siunitx}
\usepackage{color}
\usepackage[utf8]{inputenc} 
\usepackage{upgreek}
\usepackage{blindtext}
\usepackage{enumitem}
\usepackage{mathtools}
\usepackage{lineno}
\usepackage{authblk}
\usepackage{lineno}
\usepackage[labelfont=bf]{caption}
\usepackage[figurename=Figure]{caption}
\usepackage{amsmath,amssymb}

\newcommand{\ud}{\mathrm{d}}

\usepackage{float}

\begin{document}

\title{Zero-point quantum swing of magnetic couples}

\author[1,2*]{Juba Bouaziz}
\author[3]{Julen Iba\~nez-Azpiroz}
\author[1]{Filipe S. M. Guimar\~aes}
\author[1*]{Samir Lounis}
\affil{Peter Gr\"unberg Institut and Institute for Advanced Simulation, Forschungszentrum J\"ulich \& JARA, J\"ulich D-52425, Germany}
\affil[2]{Department of Physics, University of Warwick, Coventry CV4 7AL, United Kingdom}
\affil[3]{Centro de F{\'i}sica de  Materiales, Universidad del Pa{\'i}s Vasco (UPV/EHU), 20018 Donostia - San   Sebasti{\'a}n, Spain}
\affil[*]{j.bouaziz@fz-juelich.de; s.lounis@fz-juelich.de}

\maketitle


\begin{abstract}
Quantum fluctuations are ubiquitous in physics. Ranging from conventional examples like the harmonic oscillator to intricate theories on the origin of the universe, they alter virtually all aspects of matter -- including superconductivity, phase transitions and nanoscale processes.
As a rule of thumb, the smaller the object, the larger their impact.
This poses a serious challenge to modern nanotechnology, which aims total 
control via atom-by-atom engineered devices. 
In magnetic nanostructures, high stability of the magnetic signal is crucial when targeting realistic applications in information technology, \textit{e.g.} miniaturized bits.
Here, we demonstrate that zero-point spin-fluctuations are paramount in determining the fundamental magnetic exchange interactions that dictate the nature and
stability of the magnetic state.  Hinging on the fluctuation-dissipation theorem, we establish that quantum fluctuations correctly account for the large overestimation of the interactions as obtained from conventional static first-principles frameworks, filling in a crucial gap between theory and experiment~\cite{Zhou:2010,Khajetoorians:2012}.
Our analysis further reveals that zero-point spin-fluctuations tend to promote the non-collinearity and stability of chiral magnetic textures such as skyrmions -- a counter-intuitive quantum effect that inspires practical guidelines for designing disruptive nanodevices.

\end{abstract}

\maketitle

\section*{Main}
Matter is constituted by a collection of ions and a surrounding cloud of electrons. 
These microscopic entities obey quantum mechanical laws that, in addition to thermal fluctuations, involve 
intrinsic quantum fluctuations --- a direct consequence of Heisenberg's uncertainty principle. 
The presence of fluctuations can alter the collective behaviour of particles, modifying the physical properties of matter at the macroscopic level, such as the Curie temperature of magnets~\cite{Moriya:1973}.
In addition, quantum fluctuations determine the energy of the system at its lowest level, the so-called \emph{zero-point energy} that provides an extra contribution absent in the classical world. 
A notorious signature is the long known Casimir effect~\cite{Casimir:1948}, in which an attractive force emerges spontaneously among two metallic planes separated by vacuum.
But zero-point effects can emerge in a variety of contexts, including recently found light superconducting compounds like LaH$_{10}$, a ``quantum crystal'' stabilized by atomic zero-point fluctuations~\cite{errea_quantum_2020}, nuclear spin-lattice relaxation of molecular magnets~\cite{Morello:2014}, and even the internal degrees of freedom in electrical circuit components~\cite{Shahmoon:2018}.

As realized in early works~\cite{Moriya:1973}, quantum fluctuations play a particularly relevant role concerning a central property of the electron, namely its \emph{spin}. 
Known as \emph{zero-point spin-fluctuations} (ZPSF), they
represent an essential ingredient of itinerant electron magnetism and  affect a wide range of phenomena, including  high-temperature~\cite{RevModPhys.90.015009,Moriya:2003,Patrick:2006,Stepanov:2018} and iron-based~\cite{Pengcheng:2015,Christensen:2018} superconductivity, quantum phase-transitions at 0 K~\cite{Brando:2016}, and even skyrmion lattices~\cite{roldan:2015}.
ZPSF are also predicted to play a notorious role in elemental transition-metal paramagnets and ferromagnets~\cite{Wysocki:2016,Wysocki:2017} {by modifying the effective magnitude of the spin moments and possibly inducing spin} anharmonic effects~\cite{Solontsov:1994}, which can alter the magnetic stability of the ground state~\cite{Antropov:2011}.

ZPSF become increasingly important as the size of the system is decreased down to a handful of atoms, 
a regime where quantum effects prevail. 
Due to the tremendous current appeal of such low-dimensional systems in the context of information technology and storage as miniaturized magnetic bits~\cite{Gambardella:2003,Meier:2008,loth_bistability_2012}, it becomes crucial to understand and control the effect of quantum fluctuations over their magnetic properties.
In the ultimate limit of a single magnetic adatom deposited on a non-magnetic substrate, different measurements display contrasting magnetic trends depending on the probing  protocol~\cite{Gambardella:2003,Honolka:2012,Heinrich:2004,Khajetoorians:2013,Dubout:2015,Meier:2008}; while ensemble measurements based on X-ray magnetic circular dichroism report the presence of a huge magnetic anisotropy energy (MAE) that protects the magnetic moment against external interactions, local probing techniques generally find no stable magnetic signal in the very same systems. 
To resolve this apparent contradiction, first-principles theory has successfully invoked ZPSF as a mechanism that destroys the magnetic bi-stability by locally reducing the MAE barrier; the larger the fluctuations, the larger the local reduction, clarifying many of the observed trends~\cite{Ibanez:2016,Julen:2018}.

\begin{figure}
	\centering
	\includegraphics[width=1.0\textwidth]{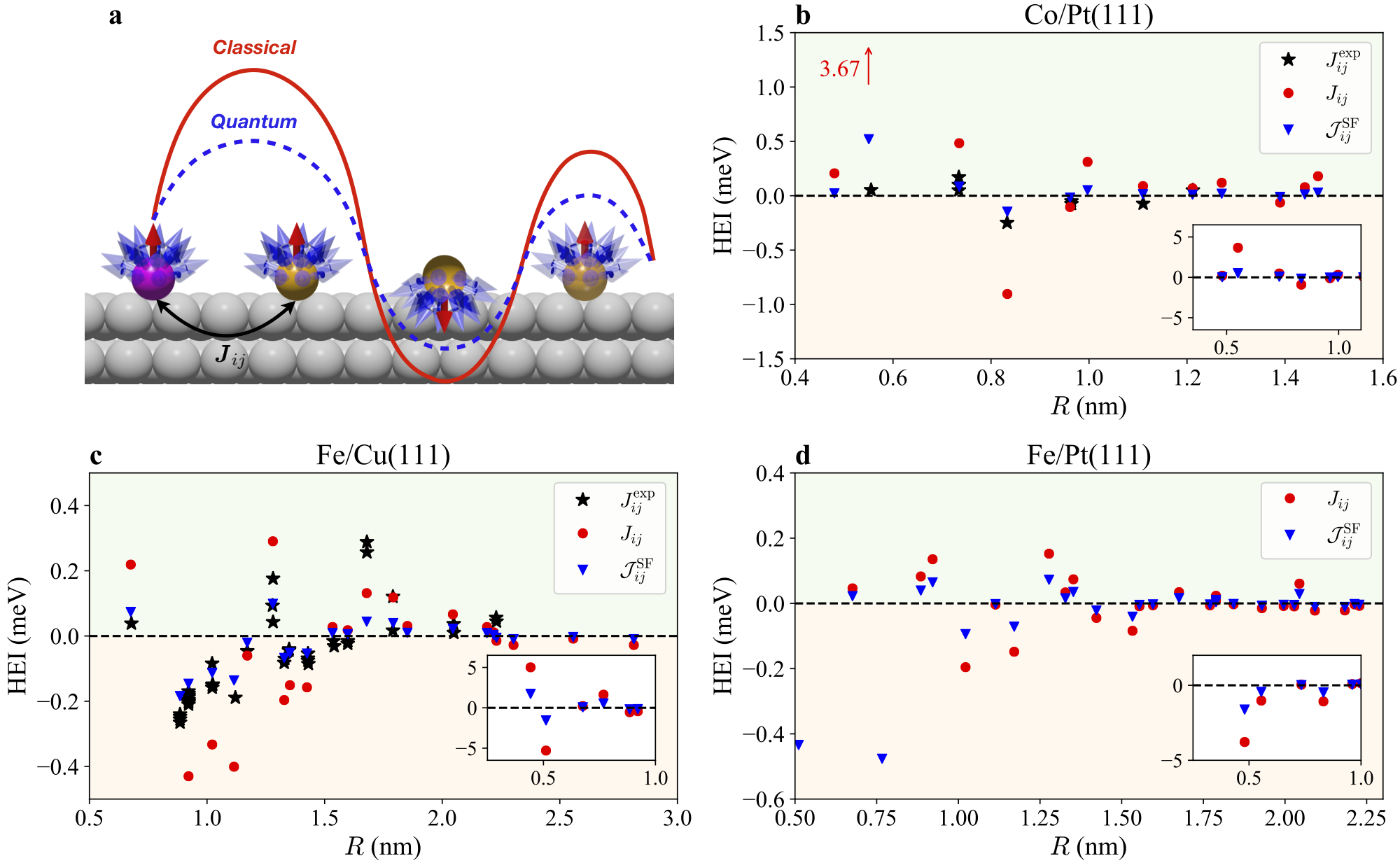}
	\caption{Distance dependency of the magnetic exchange interaction. (a) 
	Schematic depiction of the magnetic exchange interaction among two fluctuating magnetic moments as a function of interatomic distance (repeated yellow atom portrays distance-dependence). 
	The characteristic RKKY-like dependence is depicted by the oscillating curves,
	which also exhibit the strong renormalization from the 
	``Classical'' (solid red) to the ``Quantum'' (dashed blues) prediction.
	(b-d) Calculated direct exchange interaction as a function of interatomic distance of dimers Co/Pt111, Fe/Cu111 and Fe/Pt111, respectively. The bare and fluctuation-renormalized 
	Heisenberg exchange interaction (HEI) are denoted by red dots and blue triangles, respectively. In (b) and (c), the respective experimental data (black stars) from Refs.~\citenum{Zhou:2010} and~\citenum{Khajetoorians:2012} is also included with permission from Nature Publishing group. 
		}
	\label{Dimer_schema_sf}
\end{figure}

Achieving full control over the magnetic properties, however, requires understanding how quantum fluctuations affect not only  a single magnetic moment, but also  -- and crucially -- the way magnetic moments talk to each other via the \emph{magnetic exchange interaction} (MEI).
Advances in spin-polarized scanning probing techniques undergone in the last decade allowed pioneering magnetometric measurements of the most fundamental MEI between two magnetic adatoms (\textit{i.e.} a magnetic dimer)~\cite{Meier:2008,Zhou:2010,Khajetoorians:2012}.
Remarkably, first-principles calculations based on the local spin density approximation (LSDA) predict very precisely the nature of the coupling measured experimentally (\textit{i.e.} ferromagnetic [FM] or 
anti-ferromagnetic [AFM]) as a function of the interatomic distance~\cite{Zhou:2010,Khajetoorians:2012} (see Fig.~\ref{Dimer_schema_sf}a for an schematic illustration). 
However, in astonishing contrast to this success, the magnitude of the MEI obtained theoretically systematically overestimates the experimental value, with a relative error of more than 100\%, as shown in Figs.~\ref{Dimer_schema_sf}b and c.

In this work, we show that quantum fluctuations fill in the crucial gap between standard theory and experiments.
By adapting the coupling constant integral formalism~\cite{Monod:1968,Moriya:1973,Moriya:2003} to the modern
framework of time-dependent density functional theory (TD-DFT)~\cite{Samir:2010,Lounis:2011,Manuel:2015}, we demonstrate that the reduction of the MEI magnitude induced by local and non-local ZPSF results in a striking direct agreement with previous experimental results~\cite{Zhou:2010,Khajetoorians:2012}. 
In addition, our analysis reveals that anti-symmetric spin interactions of Dzyaloshinskii-Moriya-type 
are particularly robust against ZPSF,
implying that quantum fluctuations favor the emergence of chiral magnetic textures. 
These findings highlight the paramount importance of quantum effects in the study of nanoscale magnets and their future applications.

\section*{Results}
\subsection{Zero-point spin-fluctuations from first-principles.}

As the first step in our theoretical  analysis, let us obtain an expression for the zero-point energy induced by the spin-fluctuations, which can then be used to assess the impact on the MEI.
For this purpose, we make use of TD-DFT linear response theory within adiabatic LSDA, which represents one of the most powerful tools for analyzing dynamical properties of spins via \textit{ab-initio} methods~\cite{Samir:2010,Rousseau2012}.
As a mean-field theory, however, it does not incorporate near-critical fluctuations~\cite{Halilov:2006,Ortenzi:2012}, which can be key in low-dimensional systems~\cite{Ibanez:2016,Julen:2018}.
A notable way in which ZPSF can be incorporated into the theoretical framework is by making use of the so-called \emph{fluctuation-dissipation theorem}, a fundamental relation that gives access
to the {magnitude} of the local fluctuation of the magnetic moment~\cite{Moriya:2003,Ibanez:2016}:
\begin{equation}
\xi_{i,\pm}^{2} = \frac{1}{\pi} \int_{-\infty}^{+\infty}\text{Im}\,\chi^{+-}_{i}(\omega)\,\text{sgn}(\omega)\,d\omega\quad,
\label{eq:FD}
\end{equation}
with $\chi^{+-}_{i}(\omega)$ representing spatial average of the 
interacting transverse-susceptibility of a magnetic atom at site $i$~\cite{Lounis:2011,Manuel:2015}, evaluated using the Dyson-like equation 
$\underline{\chi}^{+-}(\omega) = \underline{\chi}^{+-}_{0}(\omega) + \underline{\chi}^{+-}_{0}(\omega)\,\underline{\mathcal{K}}\,\underline{\chi}^{+-}(\omega)$. 
Here,
$\underline{\chi}^{+-}_{0}(\omega)$ and $\underline{\mathcal{K}}$ represent the Kohn-Sham susceptibility, describing electron-hole excitations, and the exchange-correlation kernel, respectively; the underline bar is used to denote the tensorial character of objects.

Going one step further, we obtain the zero-point energy
associated to the ZPSF by combining the modern TD-DFT framework with 
the coupling constant integral method~\cite{Monod:1968,Moriya:2003,Moriya:2006}, giving rise to the first important relation
\begin{equation}
\begin{split}
\mathcal{E}_{\pm} 
& = -\frac{1}{2\pi}\,\text{Im}\,\text{Tr}\int_{-\infty}^{+\infty}
\left[\underline{\chi}^{+-}_{0}(\omega)\,\underline{\mathcal{K}}+\text{ln}\left(\underline{1}- 
\underline{\chi}^{+-}_{0}(\omega)\,\underline{\mathcal{K}}\right)\right]\,\text{sgn}(\omega)\,\text{d}\omega\quad,\\ 
\end{split}
\label{RPA_energies_multi_sites}
\end{equation}
where the trace runs over the number of magnetic units present in the system.

The fluctuation-corrected band energy of the system $(\mathcal{E}_\text{SF})$ is the sum of the LSDA band energy $(\mathcal{E}_\text{b})$ and $\mathcal{E}_{\pm}$ [Eq.~\eqref{RPA_energies_multi_sites}]. 
Then, assuming an extended Heisenberg Hamiltonian of the form $\mathcal{E}_{\text{SF}} = -\frac{1}{2}\sum_{i\ne j}\vec{e}_{i}\underline{\mathcal{J}}^\text{SF}_{ij}\vec{e}_{j} + \mathcal{E}_{\text{MAE}}$ ($\mathcal{E}_{\text{MAE}}$ denotes the local contribution from the onsite MAE),
and relying on the magnetic force theorem together with the infinitesimal rotational method~\cite{Liechtenstein:1986,Udvardi:2003,Ebert:2009},
the elements of the MEI tensor are determined from the curvature of $\mathcal{E}_\text{SF}$
as
\begin{equation}
\mathcal{J}^{\text{SF},\alpha\beta}_{ij} \simeq - \frac{\partial^2\mathcal{E}_\text{b}}{\partial e^{\alpha}_{i}\partial e^{\beta}_{j}}
+
\frac{1}{2\pi}\,\text{Im}\,\text{Tr}\int_{-\infty}^{+\infty}
\underline{\chi}^{+-}(\omega)
\,\frac{\partial^2\left[\underline{\chi}^{+-}(\omega)\right]^{-1}}{\partial e^{\alpha}_{i}\,\partial e^{\beta}_{j}}\,\text{sgn}(\omega)\,\text{d}\omega\quad,
\label{fluct_exchange}
\end{equation}
where $e^{\alpha}_{i}$ denotes the 
transverse component $\alpha$ of the vector $\vec{e}_{i}$ defining 
the orientation of the magnetic moment at site $i$.
As a final step, we derive a simple expression for
the renormalized MEI in terms of the fluctuating moments by
mapping the \textit{ab-initio} $\chi^{+-}(\omega)$ to the one obtained from the Landau-Lifshitz-Gilbert (LLG) model. 
This allows the identification of the analytical dependence between the dynamical susceptibility and the MEI, giving rise to
\begin{equation}
\mathcal{J}^{\text{SF},\alpha\beta}_{ij} = J^{\alpha\beta}_{ij}\left[1-\frac{J^{\alpha\beta}_{ji}}{J^{\alpha\beta}_{ij}}\left(\frac{\xi^{2}_{i,\pm}}{M^2_{i}}+\frac{\xi^{2}_{j,\pm}}{M^2_{j}}+\frac{\xi^{2}_{ij,\pm}}{M_iM_j}\right)\right]\quad.
\label{Spin_fluctuations_central}
\end{equation}
Above, 
$J^{\alpha\beta}_{ij}=-{\partial^2\mathcal{E}_\text{b}}/{\partial e^{\alpha}_{i}\partial e^{\beta}_{j}}$ denotes an element of the bare magnetic exchange tensor, and $\xi^{2}_{ij,\pm}$ is a non-local ZPSF contribution;
its explicit expression, together with the detailed derivation of all previous relations, is provided in Supplementary Notes 1 to 4.

Eq.~\eqref{Spin_fluctuations_central} represents a central result of 
the present work, as it provides a quantitative expression
showing how the standard tensor of MEI is 
renormalized by quantum spin-fluctuations.
As a general and most important trend, 
this equation shows in a clear fashion that
fluctuations systematically reduce the predominant symmetric 
component of the tensor of MEI, 
given that the hierarchy 
${|\xi_{i,\pm}|}<|M_{i}|$ holds in general;
we will refer to this important contribution  ($\mathcal{J}^{\alpha\alpha}_{ij}$) as the
\emph{Heisenberg exchange interaction} (HEI) component. 
On closer inspection, Eq.~\eqref{Spin_fluctuations_central}
further reveals
that the anti-symmetric piece of the tensor ($\mathcal{J}^{\text{SF},\alpha\beta}_{ij}$ with $\alpha \ne \beta$) giving rise to the Dzyaloshinskii-Moriya interaction (DMI)
\emph{increases} upon the action of ZPSF due to its antisymmetric nature, \textit{i.e.}, 
the property 
${J}^{\alpha\beta}_{ij} = - {J}^{\alpha\beta}_{ji}$. 
The combination of these remarkable features may have profound implications for
the stability, wavelength and shape of intensively-studied chiral spin-textures, 
such as spirals, skyrmions and anti-skyrmions~\cite{Fert:2017},
which are strongly dependent on the ratio of DMI and HEI.

\subsection{Quantum corrections against experimental evidence.}

With the purpose of connecting the developed first-principles framework to the microscopic experimental designs of Refs.~\citenum{Zhou:2010,Khajetoorians:2012}, we investigate magnetic dimers deposited on non-magnetic metallic substrates as a function of interatomic distance,
as schematically depicted in Fig.~\ref{Dimer_schema_sf}. 
This analysis exposes the oscillatory 
behaviour of the MEI, bringing into play different exchange mechanisms. 
In the short range limit, direct exchange dominates due the strong hybridization among the adatom's $d$-orbitals, while for larger distances the Ruderman–Kittel–Kasuya–Yoshida (RKKY) mechanism prevails, indirectly mediating the interaction via the conduction electrons of the metallic substrate~\cite{Ruderman:1954,Kasuya:1956,Yosida:1957}.

We consider Co and Fe dimers deposited on the fcc stacking sites of the Pt(111) and 
Cu(111) surfaces, respectively; these systems were assessed experimentally 
-- via STM measurements --
and theoretically -- via bare LSDA calculations --
in Refs.~\citenum{Zhou:2010,Khajetoorians:2012}. 
In addition, we further investigate an fcc-stacked Fe dimer on Pt(111) in order to provide theoretical predictions
on the fluctuation-renormalized MEI that can be tested in future STM experiments. 
The details for the computational setups are provided in the Methods section, while the
calculation details can be found in the Supplementary Material.
In a nutshell, we first determine the nature of the coupling (\textit{i.e.} FM or AFM) using the 
infinitesimal rotation method, while the easy axis (MAE) is resolved from the static transverse 
susceptibility~\cite{Guimar:2017}. Subsequently, the spin-excitation spectrum is computed from the 
corresponding collinear ground state configuration, in the spirit of the magnetic 
force theorem approach.

We begin by briefly describing the calculated ground-state magnetic properties of
the adatoms conforming the dimers, which evolve large magnetic moments of $2.3$ $\mu_{B}$
for Co/Pt(111), $3.3$ $\mu_{B}$ for Fe/Cu(111) and $ 3.5$ $\mu_{B}$ for Fe/Pt(111). 
In all systems, the MAE is of the order of few meV and 
favors an out-of-plane orientation.
These values are found to be very mildly affected by the interatomic distance.

\begin{figure}
	\centering
	\includegraphics[width=1.0\textwidth]{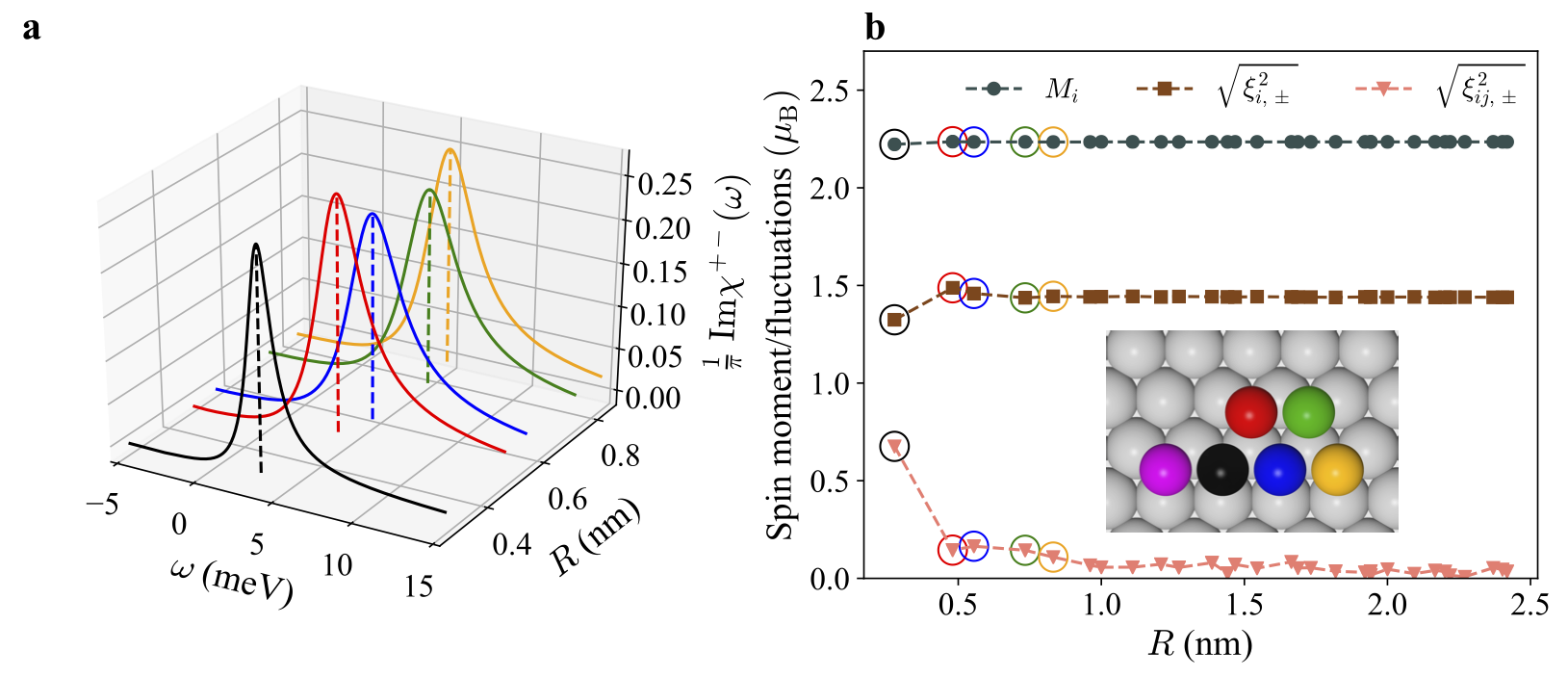}
	\caption{Spin excitations and zero-point spin-fluctuations for Co/Pt(111).
	(a) Transverse spin-excitation spectra computed at the five shortest  
	interatomic distances. The corresponding atomic configuration is depicted in the inset of
	panel (b); the magenta sphere represents a fixed atom, while the other spheres
	correspond to varying positions of the second atom, whose colors  
	correlate with the lines of $\text{Im} \chi^{+-}(\omega)$. 
	(b) Calculated magnetic moment, local and non-local zero-point spin-fluctuations are depicted by circles (grey),
	squares (brown) and triangles (pink), respectively. Results are shown as a function
	of interatomic distance; the five shortest  distances are encircled by colors that correlate with
	the atomic configuration portrayed in the inset.}
	\label{Dimer_spectra}
\end{figure}

Taking Co/Pt(111) as case study, next we illustrate the main properties of 
the spin-excitation spectrum of the dimer as a function of interatomic distance in Fig.~\ref{Dimer_spectra}a.
Our calculations reveal that the spectrum is largely dominated by a strong acoustic mode at $\sim \SI{4}{\milli\electronvolt}$.
The optical mode (not shown) is much weaker, lying above \SI{60}{\milli\electronvolt} for the shortest distance  and quickly merging with the acoustic mode as the atoms are moved far apart. 
We note that the position of the acoustic mode correlates with the magnitude of the MAE, while the broadening of the spin-excitation is induced by electron-hole excitations~\cite{Samir:2010,Manuel:2015}.
Our results evidence two clear regimes as a function of interatomic distance: the \emph{dimer-like} ($\lesssim \SI{0.7}{\nano\meter}$) and the \emph{atomic-like} ($\gtrsim \SI{0.7}{\nano\meter}$).
In the dimer-like regime, the symmetry is lowered~\cite{Guimar:2017} and the interaction between the two magnetic adatoms makes the acoustic mode highly distance-dependent, converging to the single-adatom limit (atomic-like regime) as the strength of the interaction decays.

Owing to the relation established by the 
fluctuation-dissipation theorem [\textit{c.f.} Eq.~\eqref{eq:FD}],
these two markedly different regimes are translated into the 
evolution of the local and non-local ZPSF 
depicted in Fig.~\ref{Dimer_spectra}b; 
while the size of the dominant local spin-fluctuations oscillates
between $\sim 1.3$ $\mu_{B}$ and $\sim 1.5$ $\mu_{B}$ at short distances, it
converges to a steady value of $\sim 1.4$ $\mu_{B}$ in the atomic-like regime.
In turn, the non-local ZPSF decay quickly as function of 
distance from $\sim 0.7$ $\mu_{B}$ to virtually zero, 
and are therefore only relevant in the dimer-like regime.
Notably, the calculated magnitude of the ZPSF is of the same order as the 
ground-state magnetic moment itself 
(also depicted in Fig.~\ref{Dimer_spectra}b),
a huge relative value for a purely 
quantum effect.

Having determined the evolution of the spin-excitation spectrum and ZPSF
as a function of interatomic distance  (Fig.~\ref{Dimer_spectra}),
we now assess the impact of the calculated spin-fluctuations on the 
renormalization of the HEI, namely the average of the diagonal components of $\mathcal{J}^{\text{SF},\alpha\beta}_{ij}$ in Eq.~\eqref{Spin_fluctuations_central}. Calculated results for Co/Pt(111), Fe/Cu(111) and Fe/Pt(111) 
are presented in Figs.~\ref{Dimer_schema_sf}b, c and d, respectively; for the two first systems, we additionally
include experimental STM data available from Refs.~\citenum{Zhou:2010,Khajetoorians:2012}.
The adatoms couple ferromagnetically (positive HEI) for the nearest-neighbor
distance, and display
a characteristic RKKY decaying oscillatory behavior as the interatomic distance is increased, 
with different oscillation 
periods determined by the Fermi surface 
of the substrate. 
Notably, the renormalization effects caused by quantum fluctuations
fix the disagreement between standard theory and experiments
at virtually all interatomic distances,
as demonstrated in Figs.~\ref{Dimer_schema_sf}b and c.
As revealed by our results, the main role of the fluctuations is to systematically 
reduce the magnitude of the HEI, correcting for the overestimation arising from LSDA.
We note that in Refs.~\citenum{Zhou:2010,Khajetoorians:2012}, the bare HEI was 
systematically divided by a factor equal to either 3 or 2 in order
to bring theoretical results closer to experiments.
Strikingly, including ZPSF achieves an excellent comparison without the need
of invoking \textit{a posteriori} parameters, thus proving that quantum
fluctuations are a fundamental mechanism at play in these type of 
low-dimensional magnets.

\subsection{Understanding quantum fluctuations via the LLG model.}

In order to identify the role of the fundamental factors that determine 
the behaviour of ZPSF, we resort to the 
widely used LLG model for characterizing the microscopic spin-dynamics~\cite{Gilbert:2004}.
By working out the evolution of the magnetic moments forming a dimer (see Methods section), we 
synthesize the main properties of the local and non-local ZPSF 
as a function of two central quantities. The first one consists of 
an effective MEI weighed by the magnetic moment,
$J_\text{eff} = J_{ij}\gamma/M$, with $\gamma$ the gyromagnetic ratio
(we assumed $M_1=M_2\equiv M$ for simplicity). 
We note that this quantity accounts for the distance-dependence of the system,
given that a large (small) $J_{ij}$ corresponds to a small (large) interatomic distance.
The second ingredient is the \emph{Gilbert damping} $\eta$, 
a quantity that is closely connected to the 
width of the spin-excitation peak~\cite{azpiroz:2017prb} shown in Fig.~\ref{Dimer_spectra}a
and is known to be a key player for local ZPSF, as it
quantifies the magnitude of electron-hole Stoner excitations~\cite{Ibanez:2016} (see Methods section).
In the dimers studied in this work, the calculated values 
of $\eta$ range between $\sim0.1$ for Fe/Pt(111) and $\sim0.4$ for Co/Pt(111).

\begin{figure}
	\centering
	\includegraphics[width=1.0\textwidth]{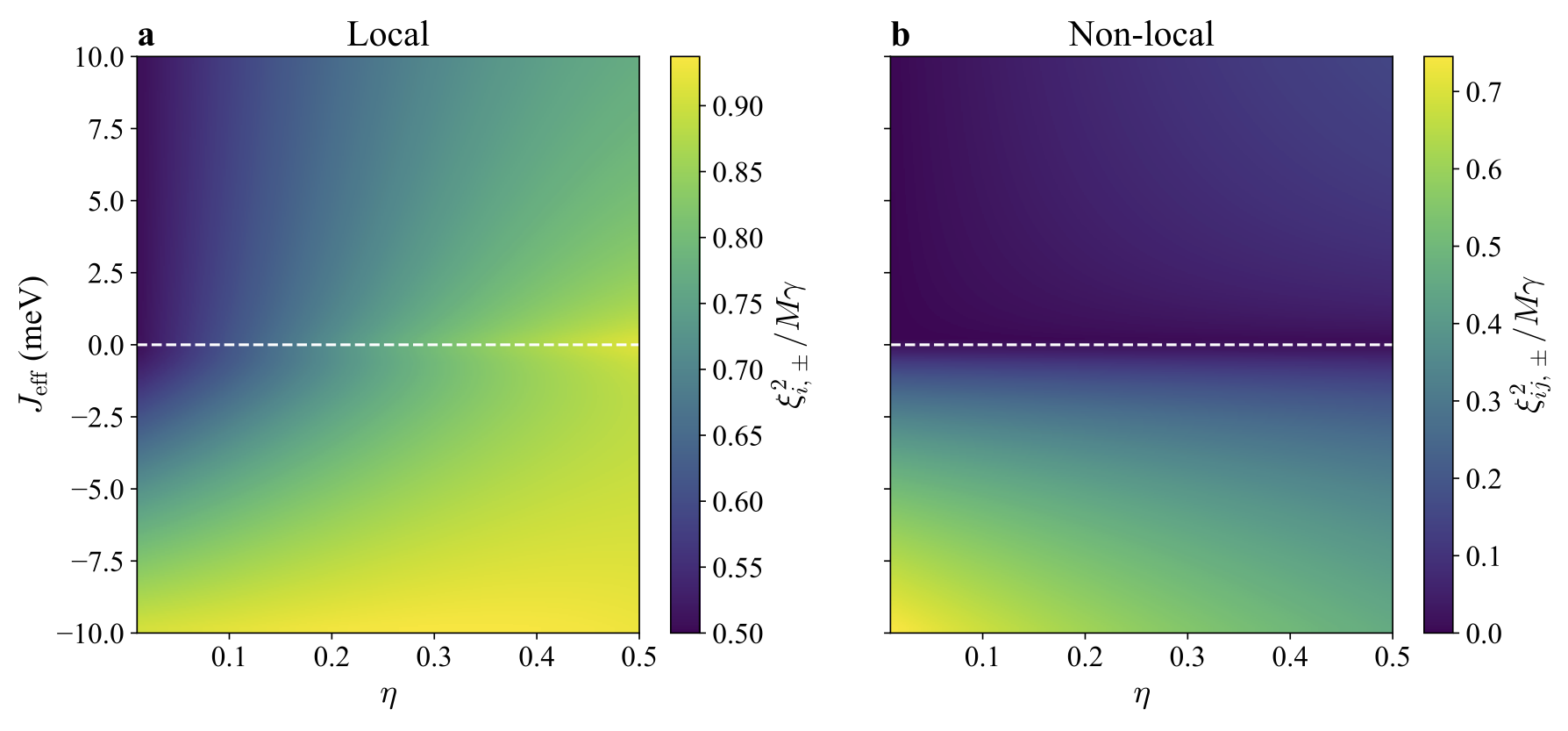}
	\caption{Colormaps of  quantum fluctuations. (a) Local and (b) non-local contributions of the fluctuation-to-magnetization ratio
	as a function of the effective exchange parameter, 
	$J_\text{eff} = J_{ij}\gamma/M$, and the Gilbert damping, $\eta$.
	Results obtained with the Landau-Lifshitz-Gilbert equation of motion (see Methods) with the  
	effective magnetic anisotropy energy (MAE), $K_\text{eff} = \text{MAE}\cdot\gamma/M$ 
	set to $\SI{1}{\milli\electronvolt}$, accounting 
	for an out-of-plane MAE in the order of magnitude of interest.
	}
	\label{LLG_spin_fluctuations_maps}
\end{figure}

Fig.~\ref{LLG_spin_fluctuations_maps} illustrates the LLG solutions for the 
local and non-local contributions to the spin-fluctuation amplitude as a function of $J_\text{eff}$ and $\eta$ in the region relevant for experiments
(we have set the MAE along the out-of-plane direction; see Methods for details).
The figure reveals valuable information on the nature of the ZPSF.
First and foremost, quantum fluctuations induce a much larger impact on the AFM regime $(J_\text{eff}<0)$ as compared to the FM one $(J_\text{eff}>0)$.
This result holds for both local and non-local contributions, and
comes to support the generally assumed notion whereby ferromagnets are more robust against 
quantum fluctuations than their anti-ferromagnetic counterpart~\cite{Guimar:2017,Holzberger:2013}.
As a second major message, Fig.~\ref{LLG_spin_fluctuations_maps}
shows that the magnitude of both
the effective MEI and
Gilbert damping play a different role depending on 
the nature of the magnetic coupling. In the FM regime, $\eta$
tends to significantly increase both local and non-local 
quantum fluctuations (in line with a previous study on the 
single-adatom case~\cite{Ibanez:2016,Julen:2018}),
whereas the effect of $J_\text{eff}$ is much milder.
In turn, the AFM regime shows a more complex pattern.
On one hand, the effect of $|J_\text{eff}|$ is significant, inducing large 
local and non-local fluctuations as it increases;
as for the Gilbert damping, it induces larger
local fluctuations but reduces the non-local ones, resulting in a competing mechanism.
As an important remark, we note that for $J_\text{eff}\simeq 0$, the non-local ZPSF vanish
both in the FM and AFM regions independently of $\eta$, a feature that explains the quick fall of 
$\xi^{2}_{ij,\pm}$ observed 
in the \textit{ab-initio} results of Fig.~\ref{Dimer_spectra}b.

In order to extract a final lesson from the LLG analysis, 
let us focus on the limit of zero damping, \textit{i.e.} $\eta\rightarrow 0$. 
This allows the identification of the intrinsic local ZPSF,
which are predominant at large interatomic distances. 
In this regime, we have $\xi_{\pm,i}^{2} \simeq \gamma M_{i}/2$, which is the lower-limit of the ZPSF (see Fig.~\ref{LLG_spin_fluctuations_maps}) and provides a poor man's approach for the renormalization of the HEI components from Eq.~\eqref{Spin_fluctuations_central}:
\begin{equation}
\mathcal{J}^{\text{SF},\alpha\alpha}_{ij} = J^{\alpha\alpha}_{ij}\left(1-\dfrac{\gamma}{2}\dfrac{M_i+M_j}{M_{i}M_{j}}\right)\quad.
\label{eq:poor-mans}
\end{equation}
Notably, this simple expression can be readily incorporated into the conventional 
DFT framework when computing the MEI with virtually no added cost, 
given that it only involves ground-state magnetic moments.
Note also that Eq.~\eqref{eq:poor-mans} provides a sensible macroscopic limit, 
whereby the renormalization due to quantum fluctuations 
tends to vanish for large values of the magnetic moment.

\section*{Discussion}

Besides shedding light into the features of the \textit{ab-initio} calculations, the parameter-space map built in Fig.~\ref{LLG_spin_fluctuations_maps} provides a simple recipe to design 
robust collinear nanomagnets.
As its central prediction, an optimal shield against quantum spin-fluctuations can be 
achieved by a combination of large ferromagnetic HEI coupling, relatively low Gilbert damping, (\textit{i.e.} low electronic hybridization) and a strong out-of-plane MAE.
On the other hand -- and in remarkable contrast to the direct-exchange mechanism --, ZPSF play in favour of the anti-symmetric nature of the DMI,
as proven by  Eq.~\eqref{Spin_fluctuations_central} 
(see also Supplementary Note 3).
This finding provides yet another major prediction, namely that quantum spin-fluctuations \emph{enhance} the non-collinearity and the stability of chiral magnetic textures such as skyrmions, governed by the DMI-to-HEI ratio that \emph{increases} upon the action of ZPSF. 
This counter-intuitive quantum effect appears to be particularly relevant 
for the ongoing miniaturization 
process of skyrmions, whose characteristic length 
scale is approaching the size of the lattice constant and classical predictions
start to break down~\cite{sotnikov2020quantum}.

Quantum spin-fluctuations thus prove to be a source of novel
physics in the realm of nanomagnets. Their crucial role for the accurate description
of the fundamental magnetic coupling mechanism has been identified, 
making them a central player for the study of nanoscale 
magnetic systems. 
For this purpose, we have provided a simple prescription 
for quantitatively assessing the impact of quantum spin-fluctuations
in the renormalization of magnetic exchange-parameters [c.f. Eq.~\eqref{eq:poor-mans}],
an expression that can be readily applied to the widespread DFT framework. 
Furthermore, the zero-point effects studied here are expected to induce an
impact on macroscopic properties as well, including the exchange-stiffness and  
the Curie temperature of magnets. 
We believe that these future steps, alongside with the ones already
taken in this work, will pave the way for the quest of designing 
nano-devices for information technologies based on 
classical or quantum bits (\textit{e.g.},
RKKY-based~\cite{Khajetoorians1062,hermenau_gateway_2017,Hermenau2019}), 
both of which hinge on quantum fluctuations.

\begin{methods}
\subsection{DFT computational details.}
The first-principles simulations have been carried out using the scalar-relativistic 
all-electron Korringa-Kohn-Rostoker (KKR) Green function method, 
including the spin-orbit interaction self-consistently~\cite{Papanikolaou:2002,Bauer:2014}. 
An angular momentum cutoff 
of   $l_{\text{max}} = 3$ and a k-mesh of $600 \times 600$ 
have been used for the construction of the Green functions in real space. 
The magnetic impurities have been embedded into a real-space impurity cluster containing 
$56$ sites and $30$ substrate atoms. The magnetic excitations have been assessed in the
framework of TD-DFT~\cite{Samir:2010,Manuel:2015}, using the adiabatic LSDA for the 
exchange-correlation kernel. 
The ZPSF  have been obtained from the frequency integral of the imaginary part of 
the magnetic susceptibility, with a cutoff frequency set at $\SI{250}{\milli\electronvolt}$ (knowing that the spin-excitation energies are around few meV), 
after which a $\omega^{-2}$ 
decay is assumed~\cite{Ibanez:2016,Julen:2018}.

\subsection{Magnetic susceptibility from the Landau-Lifshitz-Gilbert equation of motion.} 
We have considered the LLG equation describing the damped precessional
motion of two magnetic moments on top of a substrate:
\begin{equation}\label{llg_methods}
\begin{split}
\frac{\ud \mathbf{M}_i}{\ud t} = -\gamma\mathbf{M}_i\times\mathbf{B}^\text{eff}_{i} + \frac{\eta}{M_i} \mathbf{M}_i\times \frac{\ud\mathbf{M}_i}{\ud t}
\end{split}\quad.
\end{equation}
The first term on the right-hand side represents the torque generated by an effective field $\mathbf{B}^\text{eff}_i =-\partial \mathcal{E}/\partial \mathbf{M}_i$, with  
$\mathcal{E}(\{\mathbf{M}_i\}) = \sum_i E_i(\mathbf{M}_i) - \mathbf{M}_1\underline{J}_{12}\mathbf{M}_2/{M^2}$
where $i=1,2$ enumerates the magnetic atoms, and
$E_i(\mathbf{M}_i)$ 
denotes the on-site MAE.
The second piece on the right hand side of Eq.~\eqref{llg_methods} models the damping process
that drives the magnetization back to equilibrium.

The low symmetry of the problem $(C_{s})$ 
dictates that the quantities involved in the LLG 
equation for a magnetic dimer 
must have tensorial form (see Supplementary Note 2). Here, for the sake of clarity, we 
consider a simpler model where the magnetic exchange is described by a constant 
$J_\text{eff} = J_{ij}\gamma/M$ and the MAE is modeled by $K_\text{eff} = \text{MAE}\cdot\gamma/M$.
Moreover, we assume identical magnetic adatoms, \textit{i.e.} $M_{i} = M_{j} = M$. 
The key information is encoded into the LLG spin-flip susceptibility:
\begin{equation}
\begin{split}
{\chi_{ij}^{+-}(\omega)} & = \frac{M\gamma}{2}\frac{J_\text{eff} +\big(2K_\text{eff}-(1+i\eta)\,\omega\big)\delta_{ij}}{(2K_\text{eff}-(1+i\eta)\,\omega)(2J_\text{eff}+2K_\text{eff}-(1+i\eta)\omega)}.
\end{split}
\label{eq:LLG_FM}
\end{equation}
%
We have numerically computed the frequency-integral of the imaginary part of the above expression 
using the trapezoidal rule, 
giving access to the magnitude of ZPSF  [c.f. \ref{eq:FD}]  in terms of 
the LLG quantities $J_\text{eff}$, $K_\text{eff}$ and $\eta$
(see Fig.~\ref{LLG_spin_fluctuations_maps}). 
\end{methods}

\begin{addendum}
\item We are grateful to Jens Wiebe and Roland Wiesendanger for sharing with us the experimental data published in Refs.~\citeonline{Zhou:2010,Khajetoorians:2012}. This work is supported by the European Research Council (ERC) under the European Union's Horizon 2020 research and innovation programme (ERC-consolidator grant 681405 — DYNASORE).
JIA acknowledges funding from the European Union's
Horizon 2020 research and innovation program under the
Marie Sklodowska-Curie Grant Agreement No. 839237.
We acknowledge the computing time granted by the JARA-HPC Vergabegremium and VSR commission on the supercomputer JURECA at Forschungszentrum Jülich~\cite{jureca}.  

\item[Authors contributions] S.L. initiated, designed and supervised the project. J.B. developed the theoretical
scheme accounting for the quantum spin fluctuations via the random phase approximation. J.B. performed the simulations and post-processed the data. All authors helped writing the manuscript, discussed the developed method and the results.

\item[Competing Interests] The authors declare that they have no competing financial interests. 

\item[Data and materials  availability] All data needed to evaluate the conclusions in the paper are present in the paper and/or the supplementary materials. Additional data related to this paper may be requested from the authors.  The KKR Green function code that supports the findings of this study is available from the corresponding author on reasonable request.

\end{addendum}

\bibliography{main.bbl}

\end{document}